# Collaborative Product and Process Model: Multiple Viewpoints Approach

Hichem M. Geryville[1], Abdelaziz Bouras[1], Yacine Ouzrout[1], Nikolaos S. Sapidis[2]

[1]*PRISMa Laboratory, University of Lyon 2, CERRAL-IUT Lumière, 160 boulevard de l'université 69676 Bron Cedex France {hichem.geryville, abdelaziz.bouras, yacine.ouzrout}@univ-lyon2.fr*

[2]*Department of Product and Systems Design Engineering, University of the Aegean, Ermoupolis-Syros 84100 Greece {sapidis}@syros.aegean.gr*

**Abstract**

The design and development of complex products invariably involves many actors who have different points of view on the problem they are addressing, the product being developed, and the process by which it is being developed. The actors' viewpoints approach was designed to provide an organisational framework in which these different perspectives/points of views, and their relationships, could be explicitly gathered and formatted (by actor activity's focus). The approach acknowledges the inevitability of multiple interpretation of product information as different views, promotes gathering of actors' interests, and encourages retrieved adequate information while providing support for integration through PLM and/or SCM collaboration. In this paper, we present our multiple viewpoints approach, and we illustrate it by an industrial example on cyclone vessel product.

## 1 Introduction

In today's competitive world, enterprises are moving towards providing better and better customer-centric products and services to improve market share and market size with continuously growing revenue, while the product itself is seen more complex than just a physical object. The efficiency and effectiveness of product lifecycle management becomes much more important. Product Lifecycle Management (PLM) is a strategic business approach that applies a consistent set of business solutions to support the collaborative creation, management, dissemination, and use of product definition information across the extended enterprise from concept to end of life – integrating people, processes, and information [Brown 2003]. Moreover, for more efficiency these companies need to integrate their business partners in the collaboration to optimize PLM processes by integrating some Supply Chain Management (SCM) elements.

Design and development of products are highly interactive complex processes where more problems are overlaying each others, and involving collaboration of hundreds of actors designing thousands of interrelated components and making millions of coupled decisions. We talk about multidisciplinary collaboration between actors of different domains related to product lifecycle. This kind of collaboration depends on exchanging and sharing adequate information on the product, related processes and business throughout the products' lifecycle. Thus, effective capture of information, and also its extraction, interpretation, sharing, and reuse become increasingly critical, especially when one considers the collaborators' *points of view*.

In multidisciplinary collaboration (MC), the actors express their interests (activity's focus) using a variety of conditions and/or representations, and follow different processes to deploy and extract those representations. Such multi-perspective product development embodies a number of characteristics that continue to be positioned in the core of organising the product development activity. These characteristics include the need for gathering of actors' interests during product lifecycle phases following their points of view. The multiple comprehension of the information of processes and products, and the need to reason analytically over multiple views, permit to understand the properties and consequences of the multiple-viewpoint definitions of the actors. In other words, the viewpoints approach must define the specific information needed by each actor following his interests on the product/process.

In this paper, we focus on proposing a definition of multiple viewpoints approach to organise and manage the exchange and sharing of product/process information between actors. The paper reviews two main viewpoint definitions (viewpoints on systems' specification, product design views). The paper describes the use of explicit relationships between viewpoints to manage information extraction on the product development. An example has been given to illustrate the proposed approach.

## 2  Related Works

It is difficult to find a general definition of "viewpoints" in the collaborative product development area; however, there are a lot of definitions related to the Requirements Engineering (RE) and Concurrent Engineering (CE).

One of the known viewpoints framework is the Reference Model Open Distributed Processing (RM-ODP) [ISO/IEC 1994]. The RM-ODP defines a holistic framework for the specification of all distributed systems. In order to deal with all aspects and complexities of such systems, the reference model defines five different abstractions – referred to as viewpoints – from which distributed systems may be modelled: *information, computational, engineering, and technology*. These viewpoints are sufficiently independent to simplify reasoning about the complete specification of the system; they are also generic and complementary. They enable different abstraction viewpoints, allowing actors to observe a system from different suitable perspectives [ISO/IEC 1997]. One of the main benefits of RM-ODP is that, as opposed to other approaches – as IEEE standard 1471 [IEEE 2000], the "4+1" view model [Kruchten 1995], or the Zachman's framework [Zachman 1997] – it provides precise definitions of a system of interrelated concepts rather than some, often imprecise, description of isolated ones.

Among ODP viewpoints, we are here interested in the information viewpoint which is related to the information modelling. An information specification defines the semantics of information and information processing in an ODP system, without having to worry about other system considerations, such as particular details of its implementation.

Ribière [Ribière 1998] considers "Viewpoint" as a polysemous word, i.e. its definition depends on the context of use. She defines a viewpoint as "*a perspective of interest from which an expert examines the knowledge base*". It is a general definition that can take several interpretations in different domains of application. She proposes an extension of the conceptual graph formalism to integrate viewpoints in the support and in the building of conceptual graphs. Where the viewpoints allow her to define the context of use and the origin actor of concept types introduced in a graph. The aim of her proposal is to define viewpoints to help knowledge representation with conceptual graphs for multi-expert knowledge acquisition and also to have an accessible and evolutive knowledge base of conceptual graphs through viewpoints.

Other viewpoints definitions have been given for Garlan [Garlan 1987] a *viewpoint can be defined as a simplifying abstraction of a complex structure… suppressing information not relevant to the current focus*, and for Easterbrook [Easterbrook 1993] a *viewpoint represents the context in which a role is performed*.

From the product views side, Hoffman [Hoffman, Joan-Ariyo, 2000] proposes a mechanism for maintaining consistent product views in a distributed product information database. In his work, a single repository called a "master model" in which all-relevant product data resides was proposed for the integration of different product information domains, while the other views of the product must be updated to maintain consistency. Thus, Bronsvoort [Bronsvoort, Noort, Van Den Berg, Hoek, 2001] proposed a multiple viewpoints feature modelling approach to allow a designer to focus on the information that is relevant for a particular product development phase. His system supports conceptual and assembly design, part detail design and part manufacturing planning by providing their own view on a product for each of these applications.

The scope of the presented works mainly deals with the integration of a domain point of view (e.g. design) for all actors, and the generation of the product views according to the actor's objective. However their limitation is that none of them integrate the interests of the actors within different product lifecycle phases, such as the extraction/retrieve of adequate information. In multidisciplinary collaboration, the actors need to integrate their points of view on the product along its lifecycle stages, where they also need to retrieve important information within different stages according to their focus.

Among the previous definitions, we think that Ribière [Ribière 1998] and Easterbrook [Easterbrook 1993] are nearing adapted to the product lifecycle collaboration area. However, we need to improve viewpoint definition by adopting the adaptation of actors' knowledge/information within product lifecycle and supply-chain collaboration.

In this paper, we propose an approach that makes connections with product, process, and organization information for a complete interaction between multidisciplinary actors. To give an appropriate definition of multiple viewpoints actors, we situate the actors under different views (product, process and collaborative supply chain organization). In the next section, we define our notion of viewpoints and its interactions with the product/process information within a collaborative supply chain organization.

## 3 Proposed Approach

In a previous study [Geryville, Ouzrout, Bouras, Sapidis, 2005], we defined a model called PPCO (Produt-Process-Collaboration-Organization information model) based, especially, on the collaborative and project options, this study is inspired from [Gzara, Rieu, Tollenaere, 2003] and [Nowak, Rose, Saint-Marc, Callot, Eyanard, Gzara, Lombard, 2004] where they propose the Product-Process-Organization model (PPO). The PPCO provides a base-level information model that is open, extensible, independent from any product development process, aiming at capturing the engineering and business context commonly shared in product development.

The PPCO model is based on the three elements stated previously: product, process, and organization.

- Product: the architecture of the product is defined not only by the decomposition of the final product into components, functions, behaviours, etc, but also by the interactions between all these components. The interactions may include well-specified interfaces and undesired or incidental interactions.

- Process: the product development process is generally a complex procedure involving information exchange across the many activities/tasks in order to execute the collaborative work. Various network-based methods have been used to map and study development processes.

- Organization: the organization structure determines who works with whom and who reports to whom. However, in supply chains organization we are particularly interested in the study of the communication patterns between the actors conducting the technical development work.

Our notion of a viewpoint must take into account that the viewpoint is an object encapsulating cross-cutting and partial knowledge about activity, process, and domain of discourse, from the perspective of a particular actor, or collaboration-team, in the product development process.

Fundamentally, viewpoints organize the knowledge of product/software development based on gathering of interests. A viewpoint, expresses the focuses of a particular actor, such as a developer or a representative of an area of interest captured by that viewpoint. Thus, a viewpoint may represent an area of interest within a project, a product, or a process, or may simply present a particular perspective expressed in a particular notation.

The degree of importance and reuse of information changes from one actor to another according to his objectives and activities related to the product. The use of this information depends on different actors' viewpoints, competencies, skills, responsibilities and interests on product information and product lifecycle phases. Several actors use the product information differently according to the specific requirements of their discipline.

### 3.1 The multiple viewpoints approach description

In a multidisciplinary collaboration (MC) context, the human dimension is important and corresponds to the different actors in a product lifecycle phases, the related knowledge dimension corresponds to the experience, competence and situation of the actors. Most of the authors agree that a viewpoint is strongly influenced by the domain area of the actor. Factors such as the field of expertise and specific technical interest play a role in this representation. Several actors see the product differently according to the constraints specific to their discipline.

Based on both definition of Garlan and Easterbrook [Garlan 1987],[Easterbrook 1993], we a define viewpoint as *a subset of information concerning the description of a product respecting the actor's focuses*. This definition is characterized by a context, which allows the restitution of the information that the actor want to use/retrieve, and a degree of importance/reuse he wants to give to this viewpoint.

In our case, the viewpoint permits:

- Simple seek of information of product/process within a supply chain context,
- Visualization of pieces of information according to a given process/activity,
- Comparison of information between viewpoints.

So, in a MC context we notice that each actor has one or more viewpoints following his activities in different phases of the product lifecycle or supply chain process. A viewpoint is characterised by four objects (or concept types): *i)* the **Actor** concept is described by the actor's information, and his situation, which defines his role through the collaboration, and his competence level on specific domain/discipline; *ii)* the **Viewpoint_Domain** concept is defined by the current actor activity in the product development process which is dependent on the actor's situation; *iii)* the **Viewpoint_Relationship** concept indicates the different interactions between actor's viewpoints, where the viewpoint information will be compared and gathered in one global information-set, and will be used for filtering actions; *iv)* the **Viewpoint_Objective** concept is described by a focus which defines the actor's objective according to it's activity objective and the viewpoint domain of the objective's activity. Figure 1 shows the different relationships between the four concepts of viewpoint.

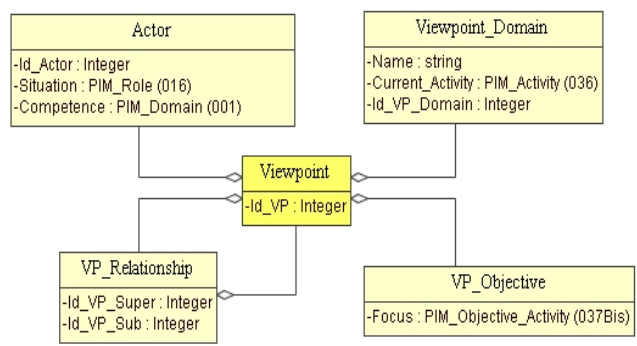

Figure 1: Viewpoint definition

In Figure 2, we consider two important parts in a viewpoint. First, the "Viewpoint_Objective" which constitutes the **focus**, second "Viewpoint_Domain" and "Actor" constitute the different **angles of view** on a same focus.

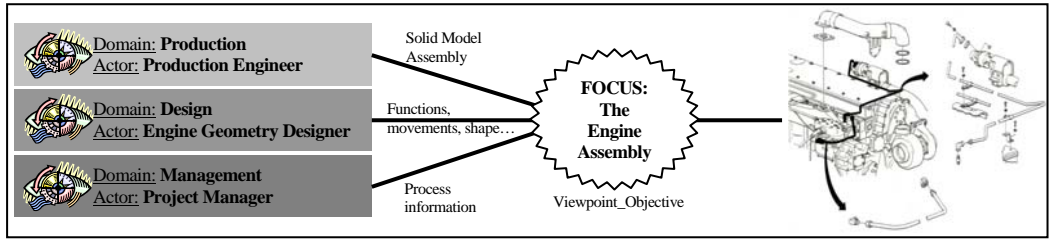

Figure 2: Example of car-engine assembly

## 3.2 Algorithm for the criteria filtering

In the table 1, we present one of the main information filtering algorithms. The filtering action consists in retrieving the appropriate information for a particular actor, according to his "viewpoint" on the product. Also it permits to combine all viewpoints and adapt the found information. However, the actors can carry out modifications, according to their rights, on the recorded information. For each modification, the actor submits his update which will be temporarily saved in the database of the PPCO model, and annotated to all actors concerned by this change. This update will take effect after approval of these concerned actors.

```
Program filtering_info_artifact (In: Artifact#, Actor#; Out: List_connexion_batch_info)
Begin
  // 1st step: Restitution of all actor's viewpoints
  Restitution_list_viewpoint(In: Actor#; Out: List_vp_actor);
  // 2nd step: Filtering of viewpoint on current artifact
  Filtering_list_vp_artifact(In: List_vp_actor, Artifact#; Out: List_vp_actor_prod);
  // 3rd step: Viewpoints classification by level of competence (decreasing classification)
  Classification_vp(In: List_vp_actor_prod; Out: List_vp_class);
  For i = 1 to size(List_vp_class) Do
    // 4th step: Restitution of information batches and the expertise level for each viewpoint
       using the focus activity and level competency
    Restitution_list_connexion_level(In: List_vp_class; Out: List_connexion_level_vp);
    // 5th step: Update of the list of information batches and its level of use
    If (i>1) Then
      Optimize_list_connexion_level(In: List_connexion_level_vp; Out: List_lev_vp_prebious);
    Else
      List_con_lev_vp_previous = List_connexion_level_vp
    End_If
  End_For
  List_connexion_batch_info = List_connexion_level_vp;
End
```

Table 1: Algorithm of information filtering by actor's viewpoints.

## 4 Example

To illustrate the proposed approach, we present an example about a closed pack cyclone vessel. Figure 3 shows the three models, product structure (or decomposition) (cf. Figure 3a), process development (cf. Figure 3b), and the development of the supply-chain organization (cf. Figure 3c). Figure 4 shows examples of two viewpoints of the same actor of the information restitution.

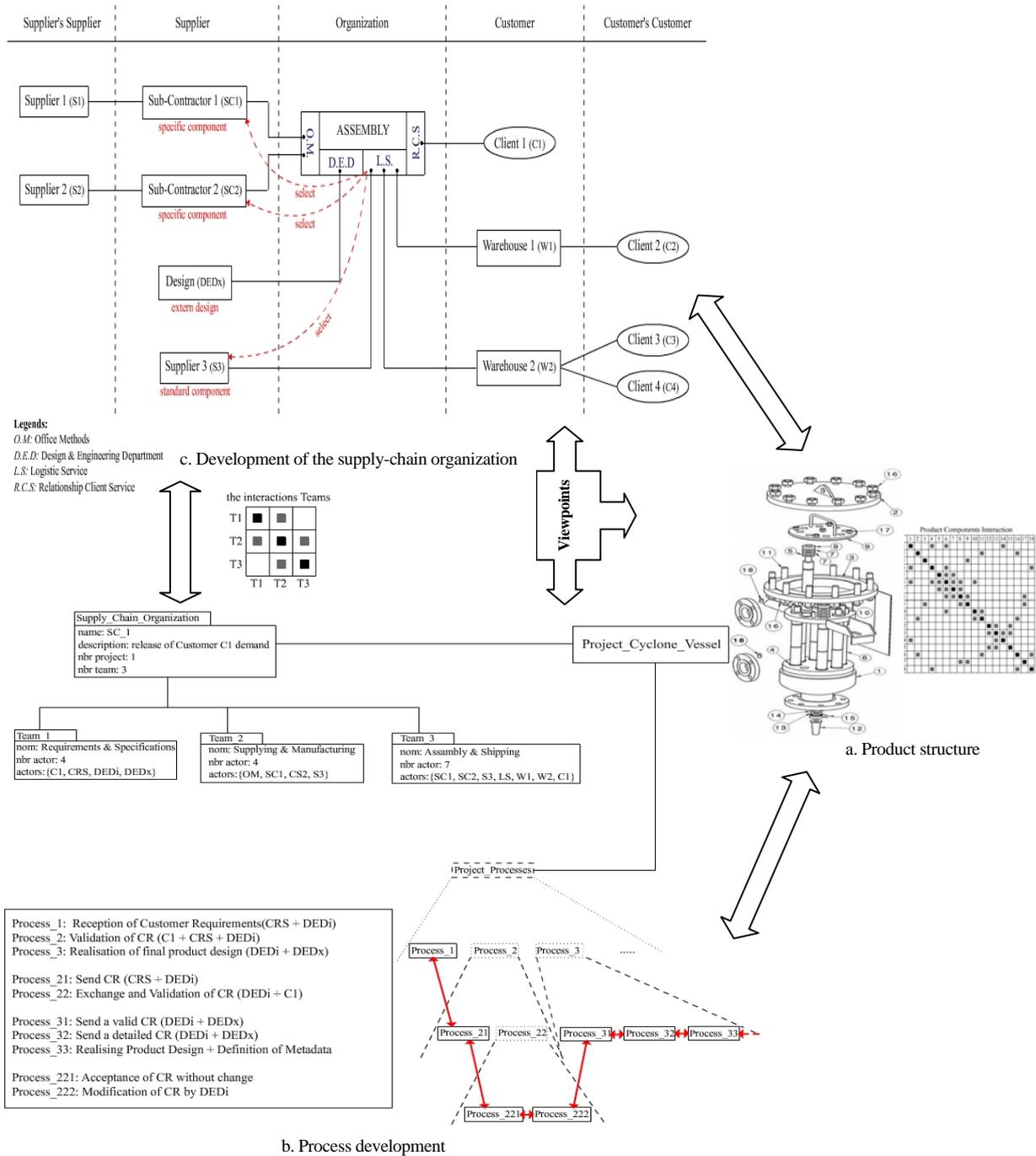

Figure 3: Examples showing interaction in the product, process, and the supply-chain models.

### 4.1 Product structure example

Figure 3a shows a model representing the decomposition of a closed pack cyclone vessel into 18 components. The product architecture and the interactions between the components are documented and detailed in this model (not presented in this paper). 38 interactions were identified according to a classification related to space, energy, material, and other information.

### 4.2 Development process example

Figure 3b shows a tree illustrating the procedure followed by a vessel manufacturer to determine the feasible layout of the customer requirements of the closed pack cyclone vessel. This is based on a digital model using CAD solid models. Interactions in this type of model represent flows of information and data between the tasks of the activities (in this example, we consider that every process has only one activity).

## 4.3 Supply-chain organization example

Figure 3c shows the decomposition of the supply-chain organization used to develop a new cyclone vessel project. The organization involves 3 teams; each team has a responsibility for a major component. The given matrix depicts the interactions across the 3 teams in terms of frequency of their required collaboration.

## 4.4 Information restitution using the viewpoint approach

Let's take the example of an actor "ActorX" integrating the supply chain organization into the team 1 as an external designer. He has two focuses on the product "cyclone vessel", the first as shape global design, and the second as mechanical design (cf. Figure 4). The actor's objectives are related respectively to the activity on geometry and mechanic tasks. To retrieve the adequate information for the actor "ActorX", we need to filter and classify the information following his interests. After this step, the framework compares the information of each viewpoint (cf. Table 2) and gives the sets of adequate information to the actor following his focuses on the product and his activities in the project. Based on the level's batch definition, the system regroups the batches with high-level hierarchy, and retrieves the information which is more adequate to the actor according to his focuses and activities [Geryville, Ouzrout, Bouras, Sapidis, 2005].

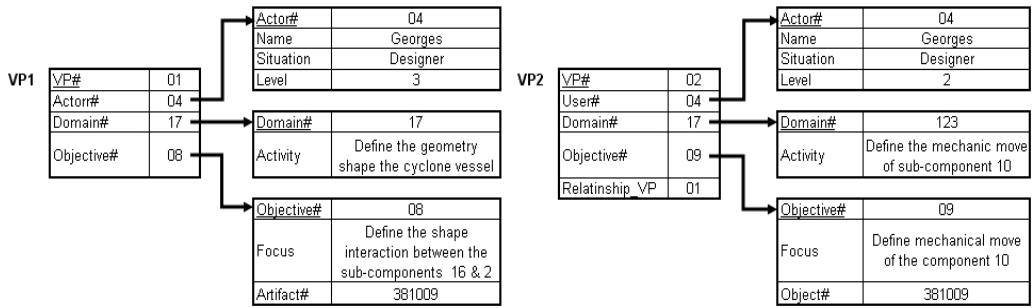

Figure 4: ActorX viewpoints

| PV# = 01 (VP1) | PV# = 02 (VP2) |
|---|---|
| Batch (Level): | Batch (Level): |
| Artifact (1) : Standard Information about the product | Mechanic (1) : All information about the mechanic application of the product (see activity of VP09 and VP08) |
| Function (2) : Different function of the final-product and sub-artifact | Artifact (2) : Standard Information about the product |
| Behavior (2) : Different behaviors of the final-product and sub-artifact in relation with their respective functions | Function (2) : Different function of the final-product and sub-artifact |
| Flows (2) : Different flows of the final-product functions | Behavior (2) : Different behaviors of the final-product and sub-artifact in relation with their respective functions |
| Geometry-Form (1) : All information about the detailed-geometry with a CAD model | Flows (3) : Different flows of the final-product functions |
| Sub-Artifact (2) : Different information about the second level of direct-component assembly | Geometry-Form (2) : Different information about the geometry with a CAD model |
| Assembly (2) : The relationship with Sub-Artifacts | Sub-Artifact (3) : Different information about the first level of direct-component assembly |
| Constraints (1): All constraints of the product (design, assembly…) | Assembly (3) The relationship with Sub-Artifacts |
| Requirements (2) : Requirements about the product and different phases of its lifecycle | Constraints (1) : All constraints design of the product |
| Group (1) : All information about actors of collaboration-team | Requirements (3) : Requirements about the product and different phases of its lifecycle |
| ... | Group (1) : All information about the actors in the group |
| | … |

Table 2: Results after the information filtering by the actor's viewpoints.

## 5 Conclusion

Multidisciplinary collaboration is a complex domain, in which all the actors need to exchange and share product and process information. In fact, product information generated by each actor is communicated to all actors in order to integrate them in a shared representation. Knowledge about actors' preferences, methods, etc. and about actors' focuses, constraints, objectives, etc. must be taken into account to manage information extraction.

In Multidisciplinary collaboration, the use of viewpoint in the structured collaborative product development shows how the viewpoint notion can provide real help in the extraction, treatment and consulting of adequate product/process information. The proposed viewpoint description and multi-level management approach aim to structure the actor's focuses in Multidisciplinary collaboration thanks to a more accurate characterization of the viewpoints, which allows an intelligent indexation of the product/process information.